\documentclass[aps,twocolumn,pre,groupedaddress,showpacs,showkeys]{revtex4}
\usepackage[T1]{fontenc}
\usepackage[latin1]{inputenc}
\usepackage{graphics}

\makeatletter

\providecommand{\LyX}{L\kern-.1667em\lower.25em\hbox{Y}\kern-.125emX\@}
\let\SF@@footnote\footnote
\def\footnote{\ifx\protect\@typeset@protect
    \expandafter\SF@@footnote
  \else
    \expandafter\SF@gobble@opt
  \fi
}
\expandafter\def\csname SF@gobble@opt \endcsname{\@ifnextchar[%]
  \SF@gobble@twobracket
  \@gobble
}
\edef\SF@gobble@opt{\noexpand\protect
  \expandafter\noexpand\csname SF@gobble@opt \endcsname}
\def\SF@gobble@twobracket[#1]#2{}

\usepackage[T1]{fontenc}
\usepackage[latin1]{inputenc}
\usepackage{graphics}

\makeatletter

\usepackage{natbib}
\usepackage{graphicx}
\usepackage{bm}
\usepackage{amsfonts}
\usepackage{amssymb}
\usepackage{amsmath}
\def\dd{\! : \!}

\makeatother

\makeatother

\begin{document}

\title{Fracture of disordered solids in compression as a critical phenomenon:\\ 
I. Statistical mechanics formalism}

\author{Renaud Toussaint} 

\email[email: ]{Renaud.Toussaint@fys.uio.no} 

\altaffiliation[Present address: ]{ Department of Physics, 
University of Oslo, P.O. Box 1048 Blindern, 0316 Oslo 3, Norway} 

\affiliation{G\'eosciences Rennes, Universit\'e de Rennes 1, 35042 Rennes Cedex, France}

\author{Steven R. Pride}
\email[email: ]{Steve.Pride@univ-rennes1.fr} 

\affiliation{G\'eosciences Rennes, Universit\'e de Rennes 1, 35042 Rennes Cedex, France}

\date{\today}

\begin{abstract}This is the first of a series of three 
articles that treats fracture localization as a critical phenomenon. 
This first article establishes     
a statistical mechanics   
based on ensemble averages when fluctuations through time play no role in defining the ensemble. 
Ensembles are obtained by dividing a huge rock  sample into many mesoscopic volumes. 
Because  rocks are a disordered  
collection of grains in cohesive contact, we expect that once shear strain is  
applied and cracks begin to arrive in the system, the mesoscopic volumes will have a wide 
distribution of different crack states. These mesoscopic volumes are the members of our ensembles. 
We determine the probability of observing a mesoscopic volume to be in a given crack state 
by maximizing Shannon's measure of the emergent crack disorder subject to constraints 
coming from the energy-balance of brittle fracture.  The laws of thermodynamics, the 
 partition function, and the quantification of temperature are obtained for such 
cracking systems. 

\end{abstract} 



\maketitle

\section{INTRODUCTION}

When rocks and other disordered-solid materials are in compression and then
have an additional deviatoric strain applied to them, small stable cracks irreversibly
appear at random throughout the material. Each time the deviatoric strain is
increased, more cracks appear. In the softening regime following peak stress,
a sample will unstably fail along a plane localized at an angle relative to
the principal-stress direction. We have accumulated evidence suggesting that
such localization is a  continuous   
phase transition. 

This is the first of three articles that develops  
a statistical
mechanics that allows the possible phase transitions in a cracking solid to be
investigated. 
Many studies have assumed that fracture is a thermally-activated process and
have used a statistical mechanics based on thermal fluctuations {[}1-5{]}. 
\nocite{SWG+91}\nocite{RK89}\nocite{WLS+91}\nocite{GP95}\nocite{Mea90}
However, our interest here is with ``brittle fracture'' in which cracks appear
irreversibly and in which thermal fluctuations play no role. For this problem,
the statistics of the fracture process is entirely due to the initial quenched
disorder in the system. 

A considerable literature has developed for so-called ``breakdown'' phenomena
in systems having quenched disorder and zero temperature {[}6-23{]}. 
\nocite{BH98}\nocite{DAH89}\nocite{DLB87} \nocite{AS93a}\nocite{SA96} 
\nocite{fracture-o:1989}\nocite{KHH97}\nocite{burst-aval:1994} 
\nocite{the-distri:1992}\nocite{failur-of-:1996}\nocite{MGP00} 
\nocite{DAHR89}\nocite{HHR91b}\nocite{HR90}\nocite{ZRS+97}\nocite{ZRS+99}
In particular, the burned-fuse {[}6-8{]},\nocite{BH98}\nocite{DLB87} spring-network
{[}9-11{]},\nocite{AS93a}\nocite{fracture-o:1989} and fiber-bundle {[}12-17{]} 
\nocite{KHH97}\nocite{MGP00}
analog models for fracture have all been shown to yield various types of scaling
laws prior to the point of breakdown {[}18-23{]} \nocite{DAHR89}
 \nocite{ZRS+97}\nocite{ZRS+99}. 
Our work is different  in that we directly treat the fracture
problem (not an analog model of it)  assuming that all of the statistics
is due to quenched disorder.  We obtain  
  the probability  of 
emergent damage states by maximizing Shannon's entropy subject to appropriate 
constraints. This  approach has recently been proven exact in the special 
case of fiber bundles \cite{PT02}.  

 The principal conclusion of our present theory is that at a critical-strain
point, there is  a continuous phase transition from states where cracks
are uniformly distributed to states where coherently oriented cracks are grouped
into conjugate bands. Several facts  justify classifying  
 such  band formation as  a critical phenomenon. 

First, the localization of the cracks into bands spontaneously breaks both the
rotational and translational symmetry of the material  even though
our model Hamiltonian preserves these same symmetries. 
The entropy of the material remains continuous and the
ensemble of the most probable states becomes degenerate at the localization
transition; {\em i.e.},  prior to  localization, the most probable state 
is the intact state, while  right at the
transition,    certain banded states acquire   
the same probability as the intact state.   
Further,  an autocorrelation length associated with the aspect
ratio of the emergent crack bands diverges in the approach to the critical point.
Unfortunately, quantitative laboratory measurements of how the bands of cracks
coalesce and evolve in size and shape prior to the final localization point do
not presently exist. 
We speculate in the third article of
this series on how such measurements might be performed. 

Our  explanation of localization based on the physics of interacting 
cracks is  distinct from  
the bifurcation analysis of  
Rudnicki and Rice \cite{RR75} in which localization is a  
consequence of a proposed phenomenological  elasto-plasticity law. 
Our work provides a method for obtaining such a plasticity law 
from the underlying physics.  

\section{THE PROBABILISTIC NATURE OF THE FRACTURE PROBLEM}

Rocks  are a disordered collection of grains in cohesive
contact.  The grains have varying shapes and sizes with typical grain sizes 
in the range of 10--100 \( \mu  \)m but sometimes considerably larger. The
contacts between the grains are generally weaker than the grains themselves
and have strengths and geometries that vary from one contact to the next. When
deviatoric (\emph{i.e.}, shear) strain is applied to a rock, grain contacts
begin to break. In what follows, a broken grain contact will be called a ``crack''.
Such a break is a stress-activated irreversible process. Once a grain contact
is broken, there is no significant healing that occurs. Cracks are not arriving
and disappearing due to thermal fluctuations. This fact makes our definition
of statistical ensembles quite different from that in the usual application
of statistical mechanics to molecular systems as we now go on to discuss.

\subsection{Creating a statistical ensemble}

\label{statens}
We imagine dividing a huge (formally infinite) system into mesoscopic
volumes that will be called ``mesovolumes''. Because the materials of interest
here have a wide range of grain-scale disorder, many different crack states
will emerge in the various mesovolumes once energy has been put into the system
and cracking begins. These various mesovolumes and the crack states they contain
comprise the ensembles in our theory. 

In order to be specific with our ideas, we now introduce a simple model of the
initial disorder and emergent crack states. The purpose of this special
model in the present paper is to motivate how ensembles are formed; however,  
the model Hamiltonian developed
in Paper II will be based upon it.  

In the model,  each mesovolume
is divided into \( N \) identical cells where a cell has dimensions on the order of a
grain size and where \( N \) is a large number such as \( 10^{2D} \)
or more with \( D \)  the system's dimension. 
In each cell, only a single grain contact is allowed to break. 
The local order parameter (explicitly  defined in Paper II)  characterizes 
both the orientation and length of such a broken grain contact.  In   
 the present paper, an order-parameter description is not yet necessary.  
 Prior to breaking, all cells are assumed to have the same elastic moduli.

 The quenched disorder is in  
 how the grain-contact breaking energy \( {\mathcal{E}}({\bm {x}}) \)
is distributed in the cells \( {\bm {x}} \) of a mesovolume.  We  
assume that only a fraction of the nominal grain-contact area  is actually 
cemented together, and that the degree of cementation from one contact to the 
next is random.  
Thus, the breaking energies  
 \( {\mathcal{E}}({\bm {x}})\) 
are  random variables independently sampled from a distribution $\pi({\cal E})$ 
having support on 
$[0,\Gamma d^{D-1}]$ where 
\( \Gamma  \) is the surface-energy density
of the  mineral, $d$ is the nominal linear dimension of a  grain contact, 
and $d^{D-1}$ is the grain-contact area in $D$ dimensions.  
The quenched-disorder distribution $\pi({\cal E})$ can have any assumed form.

We now define an infinite collection of distinct mesovolumes by allowing for
every conceivable way that \(  {\cal E} ({\bm {x}}) \) may be distributed 
in a mesovolume.
Putting this collection together forms the infinite rock mass whose properties
we are interested in determining. Each mesovolume so defined is a deterministic
system and upon slowly applying the same strain tensor \( {\bm \varepsilon } \)
to all the mesovolumes, each will undergo a deterministic cracking scenario
and end up in a well defined crack state. We denote each of the possible final
crack states with an index \( j \). 
A principal goal of the present paper is to obtain the occupation probabilities
\( p_{j} \) of these various crack states which are simply the fraction of
the mesovolumes in the system that are in the state \( j \). 

We can understand how the various crack states
emerge by appealing to a form of Griffith's \cite{Gri20} criterion. A cell  
will break only if the change in the elastic energy due to the break is greater
than or equal to the bond-breaking energy \( {\mathcal{E}}({\bm {x}}) \).
If \( {\textbf {C}}_{a} \) is the effective elastic-stiffness tensor of the
entire mesovolume that holds after the break occurs and if \( {\textbf {C}}_{b} \)
is the stiffness-tensor that held before the break,  
  Griffith's criterion 
can be stated 
\begin{equation}
\label{Grifeq}
\ell ^{D}\, {\bm \varepsilon }\dd ({\textbf {C}}_{b}-{\textbf {C}}_{a})\dd {\bm \varepsilon }/2>{\mathcal{E}}
\end{equation}
 where \( {\bm \varepsilon } \) is the strain tensor characterizing the entire
mesovolume at the moment of the break and \( \ell ^{D} \) is the volume of
a mesovolume. This particular statement  is an approximation
based on an assumed linear elasticity and absence of residual strain after unloading,
but a general statement will be derived in Section \ref{work}. Since the mesovolume
with an extra crack is more compliant than without it, the weakest cells will
begin to break even after the slightest of applied  strain. 

Yet an emergent crack state is not just a trivial consequence  of  
the ${\cal E} ({\bm x})$ distribution in a mesovolume.  Cracks 
aligned along bands  concentrate stress allowing  
even large barriers ${\cal E} ({\bm x})$ to be overtaken along the band.  
In the present model, this means that placing cracks along 
bands produces a larger change in the elastic moduli of the mesovolume than 
placing  cracks in more random positions.  
 Thus,  at least above 
some applied strain level, we expect the banded states to emerge as the ones that 
are significantly present in a rock system.  
Non-banded states at large strain are much more special.   
They can come only from mesovolumes 
in which the weak cells making
up the state are all surrounded by strong cells. 

A key idea here is that each mesovolume embedded in the system experiences
the same global strain tensor and, as such, has a crack state statistically 
independent from the other mesovolumes.   This is only valid so long 
as the emergent bands
of organized cracks have a dimension \( \xi  \) that is small relative to the
size \( \ell  \) of the mesovolume. Screening effects due to destructive strain 
interactions between incoherently oriented cracks cause the far-field strain
from a local crack structure to fall off with distance \( r \) even more rapidly
than the \( (\xi /r)^{D} \) fall off in an uncracked material. 
But even in the thermodynamic limit of infinite system sizes, 
the required statistical independence of the mesovolumes  
 breaks down right    
at the critical strain where 
 divergent bands of cracks become important.  The conclusion is that although 
our ensemble-based statistics is valid in the approach to localization, it is 
incapable of describing the post-localization physics.

\subsection{Macroscopic observables}

\label{macroobs}
In the laboratory experiments to which we apply our theory,
a sample is immersed in a reservoir from which  either uniform 
stress or strain conditions can be  applied to the sample's exterior 
surface $\partial \Omega$. 
The macroscopic strain tensor ${\bm \varepsilon}$ is defined in 
terms of the 
 displacement \( {\textbf {u}} \)
at points on \( \partial \Omega  \) as  
\begin{equation}
{\bm \varepsilon }=\frac{1}{L^{D}}\int _{\partial \Omega }{\textbf {n}}{\textbf {u}}\, dS
\end{equation}
 where \( {\textbf {n}} \) is the outward normal to the sample's surface and
\( L^{D} \) is the volume of the sample in \( D \) dimensions. This definition
of deformation thus corresponds to the volume average of the local deformation
tensor \( \nabla {\textbf {u}}({\bm {x}}) \) defined at interior points
\( {\bm {x}} \) of the sample. It will soon be shown to be  
 conjugate to the macroscopic stress tensor ${\bm \tau}$ in the expression
for the work carried out on the sample.  If strain (rather 
than stress) is the control variable,   the 
displacements at points ${\bm x}$  of the external surface $\partial \Omega$ are given by 
${\bf u} = {\bm x} \cdot {\bm \varepsilon}$.  

\begin{figure}
{\par\centering \resizebox*{7.5cm}{5.5cm}{\includegraphics{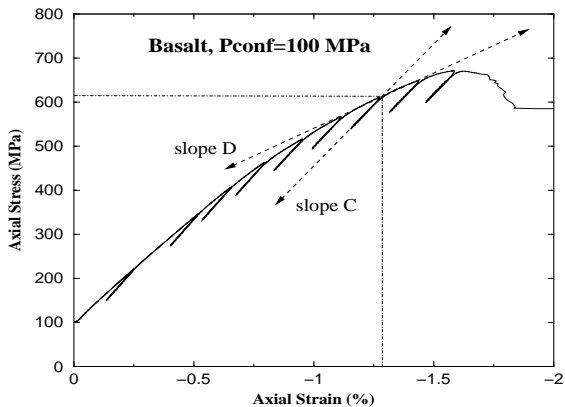}} \par}
\caption{Stress-strain data courtesy of David Lockner of the USGS Menlo Park. The slope
measured upon loading a sample is defined by \protect\protect\( {\textbf {D}}\protect \protect \)
while that measured upon unloading and/or reloading the sample is defined by
\protect\protect\( {\textbf {C}}\protect \protect \). }
\label{Lockner}
\end{figure}

As shown in Fig.\ \ref{Lockner}, a typical compression experiment starts with
the sample in a pure hydrostatic pressure state and then systematically increases
the deformation in the axial direction, keeping the radial ``confining'' pressure
\( p_{c} \) constant. Other  ways of controlling the radial stress during
the experiment are to keep a constant ratio between axial and radial stress,
or to impose a constant radial deformation.  So long as the confining pressure 
does not become so large as to induce  a brittle-to-ductile transition 
\cite{Pat78}, these various experiments all result in the same type of 
 localized structure at large axial strains.   
When  axial strain is monotonically increased, cracks arrive at each
strain increment and the deformation and stress changes are related as 
\begin{equation}
d{\bm \tau }=\frac{d{\bm \tau }}{d{\bm \varepsilon }}\dd d{\bm \varepsilon }
={\textbf {D}}\dd d{\bm \varepsilon }
\end{equation}
 where the fourth-order tensor \( {\textbf {D}} \) is called the tangent-stiffness
tensor. This tensor defines the slopes between the various stress and strain
components as the sample is being loaded and is an experimental observable. 

If at some point in the stress history the axial pressure is reduced, we follow
a different deformation path as seen in the figure due to the fact that no new
cracks are created. Such an unloading experiment defines the elastic (or secant)
stiffness tensor \( {\textbf {C}} \)
\begin{equation}
d{{\bm \tau }}={\textbf {C}}\dd d{\bm \varepsilon }.
\end{equation}
We model the unloading/reloading paths as being entirely reversible and in so doing 
 neglect the small hysteresis due to  
friction along the opened cracks. 

In order to distinguish loading paths (with crack creation) from unloading paths 
(without crack creation),  all  
 properties  are explicitly taken  to depend on two strain variables; namely, 
 the maximum  strain \( {\bm \varepsilon }_m \) having been applied to a sample, 
and the current strain  
 \( {\bm \varepsilon } \) that is different than the maximum only if the sample
has been subsequently unloaded. 
Note that even if \( {\bm \varepsilon }\) and \( {\bm \varepsilon }_m \)
are written as tensors, they each correspond to only one scalar
degree of freedom along the loading/unloading paths, since the radial components
can always be  expressed in terms  of the axial components  via the type of radial 
control employed 
({\em e.g.},  \( p_{c} = \rm{const} \)  
in a  standard  
triaxial test).  

The stress tensor ${\bm \tau}$ in the theory  corresponds to the volume average 
of the local stress tensor ${\bf T}({\bm x})$ that satisfies 
$\nabla \cdot {\bf T}(\bm x) = 0$  at interior points ${\bm x}$; {\em i.e.}, 
 \( {\bm \tau }=L^{-D}\int _{\Omega }{\textbf {T}}({\bm {x}})\, dV \) and 
is  a function of the current and maximum  strains  ${\bm \tau} = 
{\bm \tau} ({\bm \varepsilon}, {\bm \varepsilon}_m)$ as shown in Fig.\ 1.  
By averaging the elastostatic identity \( \nabla \cdot ({\textbf {T}}\, {\bm {x}})
={\textbf {T}} \) over the mesovolume we further have that 
 \( {\bm \tau }=L^{-D}\int _{\partial \Omega }{\bf n} \cdot {\textbf {T}}{\bm {x}}\, dS \).

The work density \( dU \) performed on the sample when there is an increment
in strain \( d{\bm \varepsilon } \) is in both cases of loading and unloading 
\begin{eqnarray}
dU & = & \frac{1}{L^{D}}\int _{\partial \Omega }{\textbf {n}}\cdot 
{\textbf {T}}\cdot d{\textbf {u}}\, \, dS\label{work1} \\
 & = & {\bm \tau }\dd d{\bm \varepsilon }.\label{work2} 
\end{eqnarray}
To obtain Eq.\ (\ref{work2}) from (\ref{work1}), we have written the 
controlled displacements on a sample's surface as $d{\bf u} = {\bm x} \cdot 
d{\bm \varepsilon}$ where the strain increment $d{\bm \varepsilon}$ is uniform over $\partial 
\Omega$.   
 Thus, \( dU \) corresponds to the volume average of the local
work density \( {\textbf {T}}({\bm {x}})\dd d\nabla {\textbf {u}}({\bm {x}}) \).

The total energy \( U \) per unit sample volume that goes into the sample during
the loading up to a  maximum  strain-tensor \( {\bm \varepsilon _{m}} \) is then
\begin{equation}
U({\bm \varepsilon _{m}})=\int _{\bm \varepsilon _{0}}^{\bm \varepsilon _{m}}{\bm \tau }
({\bm \varepsilon }',{\bm \varepsilon }')\dd d{\bm \varepsilon }'.
\end{equation}
 where \( {\bm \varepsilon _{0}} \) is the strain associated with the initial
isotropic stress. 
If after loading to 
\( {\bm \varepsilon }_{m} \), the sample is unloaded back to a current strain of 
\( {\bm \varepsilon } \), we have the general expression  
\begin{equation}
U({\bm \varepsilon },{\bm \varepsilon }_{m})=U({\bm \varepsilon _{m}})
+\int _{\bm \varepsilon _{m}}^{\bm \varepsilon }{\bm \tau }(
{\bm \varepsilon }',{\bm \varepsilon }_{m})\dd d{\bm \varepsilon }'.
\end{equation}
If the sample is unloaded back to the initial stress, corresponding to a possibly
non-zero residual strain \( \varepsilon^{\rm res} \), a last experimental observable
is the energy \( Q({\bm \varepsilon }_{m})=U({\bm \varepsilon }^{\rm res},{\bm \varepsilon }_{m}) \)
(per unit sample volume) that went into crack creation and that is 
lost during the loading process .

\subsection{Ergodic hypothesis}

We have shown above that the experimentally-measurable variables of energy density
\( U \), deformation \( {\bm \varepsilon } \), and applied stress \( {\bm \tau } \)
correspond to volume averages of each field throughout a system. Our ergodic
hypothesis amounts to assuming that the systems we work with are sufficiently
large that such volume averages can be replaced by ensemble averages 
\begin{equation}
\label{ensaverage}
U=\sum _{j}p_{j}E_{j},\, \, \, \, \, {\bm \varepsilon }
=\sum _{j}p_{j}{\bm \varepsilon }_{j},\, \, \, \, \, {\bm \tau }=\sum _{j}p_{j}{\bm \tau }_{j}.
\end{equation}
 Here, \( E_{j} \) is the average work per unit mesovolume required to take
an initially uncracked mesovolume from zero strain to the strain tensor \( {\bm \varepsilon }_{j} \).
A similar definition holds for \( {\bm \tau }_{j} \). In both the definition
of \( E_{j} \) and $\tau_j = dE_j/d{\bm \varepsilon}_j$, the average is over the 
initial quenched-disorder distribution.

So long as each mesovolume contains crack states that have no significant influence 
on the neighboring mesovolumes   
 (formally valid only in the thermodynamic limit), the sum 
over the collection of  mesovolumes (ensemble averaging) is equivalent 
to a volume integral over the entire system.  
In practice, we will only ever consider ensembles that have by definition 
$ {\bm \varepsilon }_{j}={\bm \varepsilon} $; however, we could equivantly 
immerse each mesovolume in a uniform stress-tensor reservoir and allow 
$ {\bm \varepsilon }_{j}$ to vary from state to state.

\section{THERMODYNAMICS OF CRACK POPULATIONS}

\subsection{Fundamental postulate}

The fracture-mechanics problem of counting how many of the initial mesovolumes
can be led to the same crack state appears to be hopelessly intractable. Fortunately,
it also appears to be unnecessary for systems containing 
 initial quenched disorder. Upon putting deviatoric strain energy into such
a system, the emergent crack states $j$ will on the one hand attempt to
mirror this quenched disorder with weakest cells breaking first; however, 
due to the energetics of the crack interactions,  
 many different types of initial mesovolumes may be led to the same crack state
which  results in  non-uniform crack-state probabilities \( p_{j} \) even if the 
quenched-disorder distribution is uniform.

We state our fundamental postulate as follows: \emph{The probability \( p_{j} \)
of observing a mesovolume to be in crack state \( j \) can be determined by
maximizing Shannon's \cite{Sha48} measure of disorder} 
\begin{equation}
\label{shannon}
S=-\sum _{j}p_{j}\ln p_{j}
\end{equation}
 \emph{subject to constraints involving the macroscopic observables that derive
from the energetics of the fracture mechanics.}  That entropy is to be maximized 
can be expected since the quenched disorder allows all states to be present 
in a sufficiently large system.   
In recent work \cite{PT02}, we have demonstrated that this postulate yields 
exact results for the special case of fiber bundles with global-load sharing.  

The constraints are what give the dimensionless function \( S \) defined
by Eq.\ (\ref{shannon}) all the thermodynamic information about our cracking
system and must explicitly involve the independent variables of $S$. 
Such independent variables are determined by establishing  the first law of 
thermodynamics 
 for a system cracking in compressive shear.

\subsection{The work of creating a crack state}
\label{work}  To obtain the first law, it is first necessary 
to define the detailed energy balance for each crack state and to understand 
how the  work \( E_{j} \) required to create  state \( j \)
depends on both the  
actual strain \( {\bm \varepsilon } \) and on the maximum-achieved strain  
\( {\bm \varepsilon}_m \).  

\subsubsection{Griffith's criterion and crack-state energy}
Consider a given mesovolume with a deterministic distribution of breaking energies
\( {\mathcal{E}}({\bm {x}}) \) assigned to each cell \( {\bm {x}} \)
of the mesovolume. Starting from a state of isotropic strain \( {\bm \varepsilon }_{0} \),
we slowly apply an additional axial deformation and monitor how one crack after
another enters the mesovolume until the final strain tensor \( {\bm \varepsilon } \)
and final crack state \( j \) are arrived at. Lets say that this state \( j \)
has a total of \( {\mathcal{N}} \) cracks associated with it.
\begin{figure}
{\par\centering \resizebox*{7.5cm}{4cm}{\includegraphics{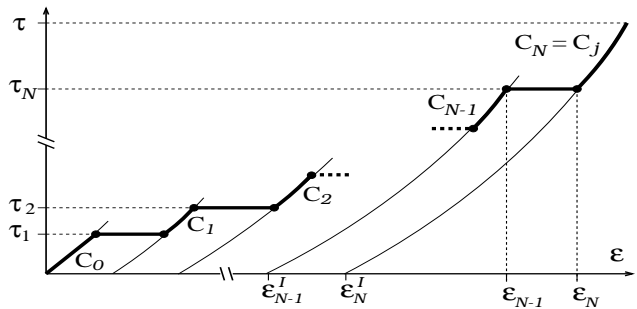}} \par}
\caption{The heavy line is the actual path followed during the steady application of
axial strain. Each vertical drop in stress corresponds to the arrival of a
crack.}
\label{crackhistory}
\end{figure}

Figure \ref{crackhistory} details the history of how the stress (and, therefore,
work) might evolve in the mesovolume as strain is applied and cracks arrive.
Initially, the mesovolume will elastically deform according to the stiffness
tensor \( {\textbf {C}}_{0} \) (no cracks yet present) until the first crack
arrives at the strain tensor \( {\bm \varepsilon }_{1} \) with an associated
drop in the mesovolume's stress. Lets say the bond-breaking energy of this first
crack was \( {\mathcal{E}}_{1} \). The mesovolume will now have a different
overall stiffness tensor \( {\textbf {C}}_{1} \) and will elastically deform
with these new moduli until the second crack arrives and so on until all \( {\mathcal{N}} \)
cracks have entered and the mesovolume has attained its final stiffness tensor
of \( {\textbf {C}}_{j}={\textbf {C}}_{\mathcal{N}} \). 
The final tensor \( {\textbf {C}}_{j} \)
depends  on both the location and orientation of these \( {\mathcal{N}} \)
cracks in addition to their number. 

At some  intermediate stage having \( n \) cracks, the stress tensor 
\( {\bm \tau }_{n}({\bm \varepsilon }) \) is defined by integrating 
\( d{\bm \tau }={\textbf {C}}_{n}({\bm \varepsilon }')\dd d{\bm \varepsilon }' \)
from \( {\bm \varepsilon }_{n}^{\rm res} \) to \( {\bm \varepsilon } \), 
where \( {\bm \varepsilon }_{n}^{\rm res} \)
is the ``residual'' deformation observed upon unloading the
sample back to zero stress as shown in the figure. We have  
\begin{equation}
{\bm \tau }_{n}({\bm \varepsilon })=
\int _{{\bm \varepsilon }_{n}^{\rm res}}^{{\bm \varepsilon }}{\textbf {C}}_{n}\dd d{\bm \varepsilon '}.
\end{equation}
 The elastic energy density corresponding to this state at deformation \( {\bm \varepsilon } \)
is similarly
\begin{equation}
E_{n}^{\text {el}}({\bm \varepsilon })=
E^{\rm res}_{n}+\int _{{\bm \varepsilon }_{n}^{\rm res}}^{{\bm \varepsilon }}{\bm \tau }_{n}
({\bm \varepsilon '})\dd d{\bm \varepsilon '}
\label{elastic1}
\end{equation}
 where \( E^{\rm res}_{n} \) represents the residual elastic energy that remains in the system
when the state with \( n \) cracks is unloaded to zero applied stress. 
These residual (zero-stress) quantities are present whenever 
plastic deformation occurs within a  grain-contact.  
After a sample elastically returns to zero applied
stress, such plastic deformation  remains  
 and, accordingly, there is an elastic stress field surrounding any crack that  experienced
plastic deformation. The  strain energy associated with such local residual
stress is what constitutes the residual energy \( E_{n}^{\rm res} \). 

When the \( n^{\textrm{th}} \)
crack arrives in a strain-controlled experiment,  there is no change in the 
strain \( {\bm \varepsilon }_{n} \) and thus no external work performed.  
However,  there is a  change in stiffness (and possibly residual strain) 
resulting in an associated stress drop \( \Delta {\bm \tau }_{n}={\bm \tau }_{n-1}
({\bm \varepsilon }_{n})-{\bm \tau }_{n}({\bm \varepsilon }_{n}) \),
and  a drop in the stored elastic energy density \( \Delta E^{\text {el}}_{n}=E_{n-1}^{\text {el}}
({\bm \varepsilon }_{n})-E_{n}^{\text {el}}({\bm \varepsilon }_{n}) \).
Energy  conservation  requires  the elastic energy reduction  to exactly 
balance the work performed in opening the crack so that  
\begin{equation}
\label{deltaEn}
-\Delta E^{\text {el}}_{n}+\frac{{\mathcal{E}}_{n}+K_{n}}{\ell ^{D}}=0
\end{equation}
 where \( {\mathcal{E}}_{n} \) is the bond-breaking work performed at 
the grain contact of the \( n^{\textrm{th}} \) crack,  \( K_{n} \) is the
energy that went into acoustic emissions when the crack arrived and/or expended
in any mode II frictional sliding or plastic deformation at the grain contact
(\( K_{n} \) is a positive ``loss'' term),  and,  as earlier, \( \ell ^{D} \)
is the volume of a mesovolume. Because \( K_{n} \) is positive, we can rewrite
Eq.\ (\ref{deltaEn}) as an inequality 
\begin{equation}
\frac{K_{n}}{\ell ^{D}}=\Delta E^{\text {el}}_{n}-\frac{{\mathcal{E}}_{n}}{\ell ^{D}}\geq 0
\end{equation}
 which is a general statement of Griffith's criterion. Upon appealing to linear
elasticity (elastic stiffnesses independent of strain level) and putting the
 residual deformation to zero (no plasticity inside the cracks), we arrive
at the convenient statement \( \ell ^{D}\, {\bm \varepsilon }_{n}\dd 
({\textbf {C}}_{n-1}-{\textbf {C}}_{n})\dd {\bm \varepsilon }_{n}/2\geq {\mathcal{E}}_{n} \)
given earlier. 

The work performed between the arrival of the \( n^{\rm th} \) and the \( (n+1)^{\rm th} \)
crack is defined  
\begin{equation}
\label{eq:intermediatework}
W_{n}=\int ^{{\bm \varepsilon }_{n+1}}_{{\bm \varepsilon }_{n}}{\bm \tau }_{n}
({\bm \varepsilon '})\dd {\bm \varepsilon '}=E_{n}^{\text {el}}({\bm \varepsilon }_{n+1})
-E_{n}^{\text {el}}({\bm \varepsilon }_{n}).
\end{equation}
Thus, the total work required to reach the final strain \( {\bm \varepsilon } \)
is the sum ({\em c.f.},  Figure \ref{crackhistory}) 
\begin{equation}
E^{p}_{j}=\sum ^{{\mathcal{N}}}_{m=0}W_{n}
\end{equation}
where by convention \( W_{{\mathcal{N}}} \) is the work performed after the
arrival of the last crack to get to the final deformation ${\bm \varepsilon}$.   
The superscript
\( p \) on \( E_{j}^{p} \) is simply indicating that this is the work for
one particular realization of the quenched disorder.   Rewriting the sum by introducing 
 Eqs.\ (\ref{eq:intermediatework}) and 
 (\ref{deltaEn}), then gives  
\begin{eqnarray}
E^{p}_{j} & = & E^{\text {el}}_{{\mathcal{N}}}({\bm \varepsilon })
-E^{\text {el}}_{0}({\bm \varepsilon }_{0})
+\sum ^{{\mathcal{N}}}_{n=1}\Delta E^{\text {el}}_{n}\nonumber \\
 & = & E^{\text {el}}_{j}({\bm \varepsilon })+\sum ^{{\mathcal{N}}}_{n=1}
\frac{{\mathcal{E}}_{n}+K_{n}}{\ell ^{D}}- E^{\text {el}}_{0}({\bm \varepsilon }_{0}) 
 \label{total2} 
\end{eqnarray}
where \( E^{\text {el}}_{0}({\bm \varepsilon }_{0}) \) is the small and
physically-unimportant amount of energy that is stored in the initial isotropic
strain field. 
Equation (\ref{total2}) is the natural statement 
that the work performed
in creating state \( j \) at strain \( {\bm \varepsilon } \) is the sum of
the elastic energy density stored in the material in the final state plus the
energy irreversibly expended during the opening of each crack. 

Both the loss term \( K_{n} \) and the residual energies \( E_{j}^{\rm res} \)  
(contained in 
\( E_{j}^{\text {el}} \)) are potentially a function of the point in
strain  history at which a grain contact actually breaks; \emph{e.g.}, most models
one might propose for  plastic deformation at a  grain contact 
are dependent on the applied stress level. However,  modeling  such plastic
processes seems uncertain at best. 
We thus  assume that at least
for those crack states significantly contributing to any phase transition (states
with lots of cracks), the stress-history dependence of \( K_{n} \) is, on 
average, negligible. Further, since the residual strain in
brittle-fracture experiments is never more than a few percent of the peak-stress 
deformation  
 and  since the essence
of the localization process does not seem to lie in \( E_{j}^{\rm res} \), 
we assume that \( E_{j}^{\rm res}\ll \sum _{n}{\mathcal{E}}_{n} \). 
With these approximations, 
   the work density \( E^{p}_{j} \)  
 depends only on the final state \( j \),  the final strain \( {\bm \varepsilon } \)
 (through \( E_{j}^{\text {el}} \)), and the  breaking energies  
 \( {\mathcal{E}}_{n} \).

The energy density \( E_{j} \)  needed  later in our  probability  
law is obtained by further averaging over the quenched-disorder in 
the breaking energies \( {\mathcal{E}}_{n} \) to give 
\begin{equation}
\label{Ej;averaged}
E_{j}= E^{\text {el}}_{j}({\bm \varepsilon }) + \gamma _{j}({\bm \varepsilon}_m) \frac{{\mathcal{N}}_{j}}{\ell ^{D}}
- E^{\text {el}}_{0}({\bm \varepsilon }_{0}).  
\end{equation}
 Here,  \( {\mathcal{N}}_{j}={\mathcal{N}} \) is the total number of
cracks in state \( j \) and  \( \gamma _{j} \) is  the average energy  
 required to  break  a single grain contact where the average is over all cells
throughout all mesovolumes led to state \( j \). 
This \( \gamma _{j} \) can be different for different final crack states. It
will also be greater at greater values of the maximum strain \( \varepsilon_{m} \)
because, according to Griffith, the cells comprising \( j \) can break at higher
energy levels when the strain is greater.   
The first term in  Eq. (\ref{Ej;averaged}) corresponds to the  purely reversible
elastic energy and therefore depends only on the actual strain state \( {\bm \varepsilon } \).

\subsubsection{Specific expression for $E_j$}
To facilitate the development in Paper II and to be more specific, 
we now use Griffith's criterion to develop  an  
expression for $E_j$ that is based   
on linear elasticity.   
When the \( n^{\textrm{th}} \)
crack arrives, the linear-elastic variant of the Griffith criterion gives that
\begin{eqnarray}
{\mathcal{E}}_{n} & < & \ell ^{D}\, {\bm \varepsilon }_{n}\dd ({\textbf {C}}_{n-1}-{\textbf {C}}_{n})\dd {\bm \varepsilon }_{n}/2\\
 & < & \ell ^{D}\, {\bm \varepsilon}_m \dd ({\textbf {C}}_{n-1}-{\textbf {C}}_{n})
\dd {\bm \varepsilon }_m/2\label{inequality2} 
\end{eqnarray}
 where as earlier \( {\bm \varepsilon }_{n} \) is the strain point on the load curve 
where the \( n^{\textrm{th}} \)
crack arrives while \( {\bm \varepsilon }_m \) is the final maximum strain level of the
experiment. 
The second inequality follows from the first since an extra crack always reduces the 
stiffness of a mesovolume.  
For any particular mesovolume in state \( j \),  the average
energy required to break a contact \( \gamma _{j}^{p} \) thus satisfies 
\begin{equation}
\label{gamma_j_p}
\gamma _{j}^{p}\equiv \frac{1}{{\mathcal{N}}_{j}}
\sum _{n=1}^{{\mathcal{N}}_{j}}{{\mathcal{E}}_{n}}<\frac{\ell ^{D}}{2{\mathcal{N}}_{j}}
{\bm \varepsilon }_m \dd \left( {\textbf {C}}_{0}-{\textbf {C}}_{j}\right) \dd {\bm \varepsilon}_m
\end{equation}
 where the right-hand side comes from summing Eq.\ (\ref{inequality2}). Since
this inequality is independent of the history, every mesovolume that is led
to state \( j \) must satisfy it. We may thus write \( \gamma _{j} \) in the
form 
\begin{equation}
\label{gamma_{j}}
\gamma _{j}=f_{j}\, \frac{\ell ^{D}}{2{\mathcal{N}}_{j}}
{\bm \varepsilon }_m\dd \left( {\textbf {C}}_{0}-{\textbf {C}}_{j}\right) \dd {\bm \varepsilon }_m
\end{equation}
 where the fraction \( f_{j} \) is  bounded as \( 0<f_{j}<1 \).
We next  demonstrate that the variation of \( f_{j} \) from one state
to the next  is so small as to be neglected altogether. 

A tighter  lower bound for \( f_{j} \) is obtained by  considering  crack states \( j \)
having \( {\mathcal{N}}_{j} \) non-interacting cracks. Since the cracks do
not interact to concentrate stress, all of the \( {\mathcal{N}}_{j} \) cells
that broke had their breaking energies somewhere in the range $0 \le {\cal E} \le 
\delta E = \ell^D {\bm \varepsilon}_m \dd \delta {\bf C} \dd {\bm \varepsilon}_m/2$ where 
$\delta {\bf C}$ is the change in the stiffness tensor  due to 
the arrival of a single non-interacting crack and $\delta E$ is the associated change in 
the elastic energy. 
Since the breaking energies are independent random variables taken from 
the   distribution $\pi({\cal E})$, we obtain  
\begin{equation}
\gamma_j  =  
\frac{\int_0^{\delta E} e \pi(e) \, de}  {\int_0^{\delta E} \pi(e) \, de} 
\label{gammagen}
\end{equation}
for non-interacting crack states $j$.

We now appeal to a specific form for the probability distribution $\pi({\cal E})$. 
Initially, our rocks are intact and it is expected that more grain contacts 
are entirely bonded (${\cal E} = \Gamma d^{D-1}$) than entirely unbonded 
(${\cal E}=0$).  We thus assume  a monotonic distribution ${\cal E}^k$ 
with $k>0$  satisfying the normalization $\int_0^{\Gamma
d^{D-1} } \pi(e) de =1$ so that  
\begin{equation}
\pi({\cal E}) = \frac{(k+1)}{\Gamma d^{D-1}}  \left(\frac{\cal E}{
\Gamma d^{D-1}}\right)^k = c \, {\cal E}^k.
\label{pi}
\end{equation}  
Using this $\pi$, the average energy required to break a contact in a non-interacting 
crack state is 
\begin{equation}
\gamma_j = \frac{k+1}{k+2} \delta E = \frac{q}{2} 
{\ell^D {\bm \varepsilon}_m \dd \delta {\bf C} \dd {\bm \varepsilon}_m} 
\label{gammaun}
\end{equation} 
where we have defined  $q = (k+1)/(k+2)$. All dependence on 
the underlying quenched-disorder distribution in our theory is confined to 
the constant $q$ which for any $k>0$ is in the range $[0.5,1]$.   

Since for non-interacting  states  
 \(  {\textbf {C}}_{0}-{\textbf {C}}_{j}
={\mathcal{N}}_{j}
 \delta {\textbf {C}} \), a  comparison of  Eqs.\ (\ref{gammaun}) and 
(\ref{gamma_{j}}) shows  that   
$f_j = q$ for all the non-interacting states.    
 For the interacting states, the prefactor \( f_{j} \) must be slightly
greater  because now stress concentration 
can allow stronger  
cells to break. It is thus concluded that for all states,  
the \( f_{j} \)  of Eq.\ (\ref{gamma_{j}}) are bounded
as \( q \le f_{j}<1 \) which when compared to how \( {\mathcal{N}}_{j} \) varies
from state to state can be considered negligible.   From here on, we simply 
take $f_j=q$ for all states.   

The essential physics for the average amount of work that goes into building up
any given crack state \( j \) is thus captured by
\begin{eqnarray}
E_{j}({\bm \varepsilon },{\bm \varepsilon }_m) & = & 
E_{j}^{R}({\bm \varepsilon }) + 
E_{j}^{I}({\bm \varepsilon }_m)
\label{hamiltonian} \\
E^{R}_{j}({\bm \varepsilon }) & = & \frac{1}{2}{\bm \varepsilon }\dd 
{\textbf {C}}_{j} \dd {\bm \varepsilon } 
\label{revenergy}\\
E_{j}^{I}({\bm \varepsilon }_m) & = & \frac{q}{2}{\bm \varepsilon }_m\dd 
\left( {\textbf {C}}_{0}-{\textbf {C}}_{j}\right) \dd {\bm \varepsilon }_m\label{irrevenergy} 
\end{eqnarray}
 where the superscripts \( R \) and \( I \) denote respectively the reversible
and irreversible part of the energy.
 The intact hydrostatic
energy \( E^{\text {el}}_{0}({\bm \varepsilon }_{0})  \) 
has been neglected since it does not involve cracks and,
therefore, cannot influence the probability of the various crack states.

\subsection{The  laws of our crack-based thermodynamics}

Using the ergodic hypothesis discussed earlier, the average energy density
in a disorded solid can be written \( U=\sum _{j}p_{j}E_{j} \). 
We are interested in how $U$ changes when increments in  ${\bm \varepsilon}$ 
and ${\bm \varepsilon}_m$  are applied to the system.  

In general, a small  increment in $U$  can be written 
\begin{equation}
\label{dU1}
dU= \sum_j E_j dp_j + \sum _{j}p_{j}dE_{j}.
\end{equation}
The first term involving the probability change  is entirely due to crack creation.
Some mesovolumes that were in less cracked states prior to the increment, are transformed 
to state $j$ during the increment,  while  mesovolumes that were in state $j$, are 
transformed to other more cracked states.  If in the increment, 
the number of mesovolumes arriving in state $j$ 
is different than the number leaving, there is a change $dp_j$ in the occupational probability 
of that state.  Such  changes are the only way to  change  the disorder in the system, so that   
\begin{equation}
\label{TdS}
\sum _{j}E_{j}dp_{j}=TdS
\end{equation}
is the work involved in changing the system's disorder via crack production.  
The  proportionality constant $T$ is formally a temperature and will be  treated  
in detail.   

Using the decomposition $E_j({\bm \varepsilon}, {\bm \varepsilon}_m)= 
E_j^R({\bm \varepsilon}) + E_j^I({\bm \varepsilon}_m)$, 
we can write  
the second term  of Eq. (\ref{dU1}) as   
\begin{equation}
\label{dU2}
\sum _{j}p_{j}dE_{j}=\sum _{j}p_{j}dE_{j}^{R}+\sum _{j}p_{j}dE^{I}_{j}.
\end{equation}
The first part is due to  purely  elastic (reversible) changes in each mesovolume and 
may be further  written  
\begin{equation}
\sum _{j}p_{j}dE_{j}^{R}={\bm \tau }\dd d{\bm \varepsilon }
\end{equation}
where ${\bm \tau}$ is the average stress tensor acting on the mesovolumes.  
This  result can be verified by appealing either to Eq.\ (\ref{revenergy}) or 
to the more general statement of  Eq.\ (\ref{elastic1}).

The second part $\sum_j p_j dE^I_j$  represents
the average work performed in creating cracks in just the final strain increment  
\(d{\bm \varepsilon }_m \). 
 Some of the initial 
mesovolumes 
led to state $j$ at maximum strain \({\bm \varepsilon}_m +  d{\bm \varepsilon }_m \) 
had all their cracks  in place before the final strain increment, while others had 
cracks arrive in the final increment.  We write 
\begin{equation}
\sum _{j}p_{j}dE^{I}_{j}={\bm g}\dd d{\bm \varepsilon }_m
\end{equation}
where the tensor ${\bm g}$ has units of stress but is quite distinct from the 
stress tensor ${\bm \tau}$.   

The ``first law''  for the rock mass is then 
\begin{equation}
\label{firstlaw}
dU={\bm \tau }\dd d{\bm \varepsilon }+{\bm g}\dd d{\bm \varepsilon }_m+TdS
\end{equation}
with the formal definitions 
\begin{equation}
{\bm \tau} = \left. \frac{\partial U}{\partial {\bm \varepsilon}} \right|_{
{\bm \varepsilon}_m, S}  , 
\, \, \, \, \, 
{\bm g} = \left. \frac{\partial U}{\partial {\bm \varepsilon}_m} \right|_{
{\bm \varepsilon}, S},  
\, \, \mbox{ and } \, \, 
{T} = \left. \frac{\partial U}{\partial {S}} \right|_{
{\bm \varepsilon},{\bm \varepsilon}_m}.  
\end{equation}
The natural variables of the fundamental function \( U \)
are \( (S,{\bm \varepsilon },{\bm \varepsilon }_m) \).   Equivalently if
\( S \) is treated as the fundamental function, 
then \( S=S(U,{\bm \varepsilon },{\bm \varepsilon }_m) \) which means 
that the constraints placed on the maximization of $S$ must involve 
$U$, ${\bm \varepsilon}$, and ${\bm \varepsilon}_m$.  

The ``second law'' of this crack-based thermodynamics is that $dS \ge 0$ 
(equal to zero only if $d{\bm \varepsilon}_m =0$ so that no cracks are created) 
while a 
 ``third law'' may be proposed by simply defining \( T=0 \)
when \( S=0 \). The system will have zero emergent disorder before cracks begin
to arrive and so our third law states that the temperature \( T \) starts at
zero and then increases in magnitude as the number of cracks in the system increases from
zero. The justification for this postulate comes \emph{a posteriori} when it
is found that in order to have zero probability for a mesovolume being in anything
but the uncracked state (\( S=0 \)), we must have that \( T=0 \).

\subsection{The  probability distribution}

To obtain the probability of observing a mesovolume to be in crack state \( j \),
we  maximize  \( S =- \sum_j p_j \ln p_j \) subject to the 
constraint that  \( \sum _{j}p_{j}=1 \), 
and to the additional  constraints that 
${\bm \varepsilon }_{j}={\bm \varepsilon }$, ${\bm \varepsilon }_{mj}={\bm \varepsilon }_m$,  
and 
$\sum _{j}p_{j}E_{j}=U$.
These constraints define our canonical ensemble. Other ensembles can be defined by considering 
other constraints involving $  {\bm \varepsilon }$, ${\bm \varepsilon }_{m}$, and $U$; 
however, since all ensembles yield identical average properties in the thermodynamic limit, 
we elect to work only with the canonical ensemble due to its  analytical     
convenience.  

This maximization problem is solved 
 using  Lagrange multipliers  to obtain the  
  Boltzmannian
\begin{equation}
\label{probaform}
p_{j}=\frac{e^{-E_{j}/T}}{Z}\mbox {\hskip 2mmwhere\hskip 2mm}Z=\sum _{j}e^{-E_{j}/T},   
\end{equation}
and where the  parameter \( T \)  is  exactly the partial derivative $ \partial U/\partial S 
|_{
{\bm \varepsilon}, {\bm \varepsilon}_m}$ called ``temperature''.

\subsection{The free energy and its derivatives}

Any equilibrium physical property that depends on the distribution of cracks
throughout the system can be obtained from the partition function \( Z \) given
by Eq.\ (\ref{probaform}). 

To do so,  a thermodynamic potential \( F \) called
the free-energy density is introduced  that is related to \( Z \) by 
\begin{equation}
\label{free}
F({\bm \varepsilon}, {\bm \varepsilon}_m,T)=-T\, 
{\ln Z({\bm \varepsilon}, {\bm \varepsilon}_m,T)}.
\end{equation}
 This potential \( F \) is the Legendre transform with respect to \( S \)
of the total-energy density \( U=U({\bm \varepsilon },{\bm \varepsilon }_m,S) \)
as can be seen from 
\begin{equation}
U-TS  =  \sum _{j}p_{j}E_{j}+T\sum _{j}p_{j}\ln p_{j}  
  =  -T\, {\ln Z}\sum _{j}p_{j}=F\label{Legendre} 
\end{equation}
 where we used that \( \ln p_{j}=-E_{j}/T-\ln Z \). 

When \( ({\bm \varepsilon },{\bm \varepsilon }_m,T) \) are the independent
variables, the first law can be obtained by taking the total
derivative of Eq.\ (\ref{free}) 
\begin{eqnarray}
dF & = & -T\frac{dZ}{Z}-\ln Z\, dT\nonumber \\
 & = & -T\sum _{j}\left[ -\frac{dE_{j}({\bm \varepsilon },{\bm \varepsilon }_m)}{T}+E_{j}\frac{dT}{T^{2}}\right] p_{j}-\ln Z\, dT\nonumber \\
 & = & (F-U)\frac{dT}{T}+\sum _{j}p_{j}\left[ dE^{R}_{j}({\bm \varepsilon })
+dE^{I}_{j}({\bm \varepsilon }_m)\right] \nonumber \\
 & = & -SdT+{\bm \tau }\dd d{\bm \varepsilon }+{\bm g}\dd d{\bm \varepsilon }_m\label{derF} 
\end{eqnarray}
 where we have used the definitions that 
\( {\bm \tau }_{j} = dE_j^{R}({\bm \varepsilon })/ d{\bm \varepsilon } \)
and 
\( {\bm g}_{j}=dE_{j}^{I}({\bm \varepsilon }_m )/d{\bm \varepsilon }_m \). 
% as well as   
% \( {\bm \tau} =\sum _{j}p_{j}{\bm \tau}_{j} \) and 
%${\bm g} = \sum_j p_j {\bm g}_j$.   

With $\beta=1/T$, the various thermodynamic functions are related to the 
partial derivatives of $\ln Z({\bm \varepsilon }, {\bm \varepsilon }_m, \beta)$ as     
\begin{eqnarray}
-\frac{\partial \ln Z}{\partial \beta } & = & \sum _{j}E_{j}p_{j}=U\\
-\frac{1}{\beta }\frac{\partial \ln Z}{\partial {\bm \varepsilon} } & = & 
\sum _{j}{\bm \tau} _{j}p_{j}={\bm \tau} \\
-\frac{1}{\beta }\frac{\partial \ln Z}{\partial {\bm \varepsilon}_{m}} & = & 
\sum _{j}{\bm g}_{j}p_{j}={\bm g}.  
%-\frac{\partial ^{2}\ln Z}{\partial \beta ^{2}} & = & 
%\sum _{j}(U^{2}-E_{j}^{2})p_{j} = \frac{1}{\beta} \frac{\partial S}{\partial \beta} = 
%\frac{\partial U}{\partial \beta}\\
%-\frac{1}{\beta }\frac{\partial ^{2}\ln Z}{\partial {\bm \varepsilon} \partial 
%{\bm \varepsilon}} & = & 
%\sum _{j}\frac{d{\bm \tau}_{j}}{d{\bm \varepsilon} }p_{j}
%+\beta \sum _{j} ({\bm \tau}{\bm \tau} - {\bm \tau}_j {\bm \tau_j}) p_{j}
%\\
%-\frac{1}{\beta }\frac{\partial ^{2}\ln Z}{\partial {\bm \varepsilon}_{m} \partial 
%{\bm \varepsilon}_m} & = & 
%\sum _{j}\frac{dg_{j}}{d{\bm \varepsilon}_{m}}p_{j}+\beta \sum _{j}({\bm g}{\bm g} 
%- {\bm g}_{j}{\bm g}_{j}) p_j \\
%-\frac{\partial ^{2}\! \ln Z}{\partial {\bm \varepsilon} \partial \beta } \!\!&\!\! = \!\!&\!\! 
%{\bm \tau} +\beta \sum _{j}E_{j}({\bm \tau} -{\bm \tau}_{j})p_{j} 
%= {\bm \tau} + \frac{1}{\beta} \frac{\partial S}{\partial {\bm \varepsilon}} 
%= \frac{\partial U}{\partial {\bm \varepsilon}}\\
%\!\!\!\!-\frac{\partial ^{2}\!\ln Z}{\partial {\bm \varepsilon}_{m}\partial \beta } 
%\!\!\!& \! \!\!= \!\!&\!\! 
%{\bm g}+\beta \sum _{j} \!\!E_{j}({\bm g}-{\bm g}_{j})p_{j} 
%= {\bm g} \!+ \!\frac{1}{\beta} \frac{\partial S}{\partial {\bm \varepsilon}_m} 
%\!= \!\frac{\partial U}{\partial {\bm \varepsilon}_m}\!.
\end{eqnarray}
These  results, along with $S=\ln Z + \beta U$,  are used in  Paper III.

\section{the temperature}
\label{tempstatus}
The temperature is a well-defined essential part of our quenched-disorder 
statistics.  Through the probability law $p_j= e^{-E_j/T}/Z$, 
the temperature quantifies 
the energy scale that separates probable from improbable states and how 
this energy scale evolves with strain. No other 
meaning should be read into $T$.   We now demonstrate how to exactly obtain $T$. 

\subsection{Evolution of temperature with strain}
The only way energy enters the  system is by performing work  on 
the external surface.  Thus, the general relation    
  $dU = {\bm \tau} \dd d{\bm \varepsilon}$ holds for either loading or 
unloading situations.  This previously unused fact   provides 
 a  differential equation for $T=1/\beta$    
 that permits everything 
about our system to be exactly known once an order-parameter based 
 model for $E_j({\bm \varepsilon}, 
{\bm \varepsilon}_m)$  is determined  and the functional sums 
defining $Z({\bm \varepsilon},
{\bm \varepsilon}_m, \beta) = \sum_j e^{-\beta E_j({\bm \varepsilon},
{\bm \varepsilon}_m)}$ are performed.  

The temperature and entropy only evolve along
 load paths defined by  ${\bm \varepsilon} = 
{\bm \varepsilon}_m$ and only such paths need be considered 
in what follows. 
 Using   
 \( dU={\bm \tau }\dd d{\bm \varepsilon } \), the first law 
 {[}Eq.\ (\ref{firstlaw}){]} can then be rewritten  
\begin{equation}
\label{eq:balance,irreversibles}
TdS+{\bm g}\dd d{\bm \varepsilon }=0.
\end{equation}
Since it always requires energy to break contacts, we have that 
 \( {\bm g}\dd d{\bm \varepsilon }>0 \)
and consequently \( TdS<0 \).  Furthermore, since  the entropy (disorder) 
necessarily grows during the crack-creation process (at least initially), the  
temperature of our  
system is negative (at least initially).  

The load path of a standard triaxial experiment is when 
axial strain $\varepsilon_z$ monotonically increases while the
radial confining stress $\tau_x=\tau_y = -p_c$ remains constant.        
Along this  path, all properties evolve only as a function of $\varepsilon_z$.
With $Z({\bm \varepsilon},
{\bm \varepsilon}_m, \beta)$ considered as  known, 
the  radial deformation components can be expressed in terms of the axial 
deformation  by  using the two equations 
$$
\beta p_c = \left. 
\frac{\partial \ln Z}{\partial \varepsilon_x} \right|_{
{\bm \varepsilon}_m = {\bm \varepsilon}}  
=\left. 
\frac{\partial \ln Z}{\partial \varepsilon_y} \right|_{
{\bm \varepsilon}_m = {\bm \varepsilon}}$$
to obtain the two functions 
\begin{equation}
\varepsilon_x = f_x(\beta, \varepsilon_z) 
\, \mbox{ and } \, \,  \varepsilon_y = f_y(\beta, \varepsilon_z) 
\end{equation}
 that are  valid  only along the load path.  

We now write $dU$ in two different ways.  First,  
$dU = {\bm \tau} \dd d{\bm \varepsilon}$ is evaluated along the load path to obtain 
\begin{equation}
dU= \tau_z d\varepsilon_z - p_c (df_x + df_y).
\label{dUv1}
\end{equation}
Second, we use the fact that $U=U({\beta, \bm \varepsilon}, {\bm \varepsilon}_m)$ 
to obtain 
\begin{eqnarray}
\!\!\!\!\!\!
&&dU= \frac{\partial U}{\partial \beta} d\beta +
\left(\frac{\partial U}{\partial \varepsilon_z}
+ \frac{\partial U}{\partial \varepsilon_{mz}}\right)
 d \varepsilon_z \nonumber \\
&& \mbox{ \hskip3mm }+ \!
\left(\frac{\partial U}{\partial \varepsilon_x} 
+ \frac{\partial U}{\partial \varepsilon_{mx}}\right)\! 
 d f_x + \!
\left(\frac{\partial U}{\partial \varepsilon_y} 
+ \frac{\partial U}{\partial \varepsilon_{my}}\right)\! 
 d f_y. 
\label{dUv2}
\end{eqnarray}
Upon equating Eqs.\ (\ref{dUv1}) and (\ref{dUv2}) we obtain 
a first-order non-linear differential equation for $\beta$  
\begin{equation}
a(\beta, \varepsilon_z)  \frac{d\beta}{d \varepsilon_z} + b(\beta, \varepsilon_z) = 0
\label{tempde}
\end{equation}
where $a$ and $b$ are given by 
\begin{eqnarray}
a= \frac{\partial U}{\partial \beta} &+& \left(p_c + \frac{\partial U}{\partial \varepsilon_x}
+ \frac{\partial U}{\partial \varepsilon_{mx}} \right) \frac{\partial f_x}{\partial \beta} \nonumber \\ 
&+& \left(p_c + \frac{\partial U}{\partial \varepsilon_y}
+ \frac{\partial U}{\partial \varepsilon_{my}} \right) \frac{\partial f_y}{\partial \beta}
\label{a} \\
b= -\tau_z &+& \frac{\partial U}{\partial \varepsilon_z} +
\frac{\partial U}{\partial \varepsilon_{mz}} \nonumber \\
&+&\left(p_c + \frac{\partial U}{\partial \varepsilon_x}
+ \frac{\partial U}{\partial \varepsilon_{mx}} \right) \frac{\partial f_x}{\partial \varepsilon_z} 
\nonumber \\
&+&\left(p_c + \frac{\partial U}{\partial \varepsilon_y}
+ \frac{\partial U}{\partial \varepsilon_{my}} \right) \frac{\partial f_y}{\partial \varepsilon_z}.
\label{b}
\end{eqnarray}
We are to use $\tau_z = - \beta^{-1} \partial \ln Z/\partial \varepsilon_z$ 
and $U= - \partial \ln Z/\partial \beta$ in these expressions for $a$ and $b$ once 
the function $Z({\bm \varepsilon}, {\bm \varepsilon}_m, \beta)$ has been determined.   
Furthermore,  
all partial derivatives are to be evaluated along the load curve; {\em i.e.}, at 
 $\varepsilon_{mx} = 
f_x(\beta, \varepsilon_z)$, $\varepsilon_{my} = 
f_y(\beta, \varepsilon_z)$, and $\varepsilon_{mz}=\varepsilon_z$.   

\subsection{Initial conditions}
In order to integrate Eq.\ (\ref{tempde}), 
initial conditions must be provided.  
The initial conditions of our so-called ``third law'' ({\em i.e.}, 
the intact conditions that $\beta=-\infty$ 
when $\varepsilon_z = 0$) are not well-defined for $\beta$. 
Thus, Eq.\ (\ref{tempde}) must be integrated not from the intact state, 
but from a state that contains at least a few cracks so that $\beta\neq -\infty$.

Accordingly,  we define ``one-crack'' initial conditions by considering 
the point in strain history where on average throughout the ensemble of 
mesovolumes, there is one crack 
 in each mesovolume.  If there are $N$ cells in a mesovolume, 
 the probability of any given cell to be broken  somewhere in the ensemble is   then 
 $P_1=1/N$.   
This same probability can  also be obtained from 
Griffith's criterion by  integrating the 
 quenched-disorder 
distribution  of Eq.\ (\ref{pi}) to obtain   
 $P_1 = [\delta E_1/(\Gamma d^{D-1})]^{k+1}$ where 
$\delta E_1=\ell^D {\bm \varepsilon}_1 \dd \delta {\bf C} 
\dd {\bm \varepsilon}_1/2$ is the elastic 
energy change due to a single isolated crack and where ${\bm \varepsilon}_1$ is the 
strain tensor at which on average there is a single crack in each mesovolume. 
  Thus, 
we have ${\bm \varepsilon}_1 \dd \delta {\bf C} \dd {\bm \varepsilon}_1= 
2 \Gamma d^{D-1}/(N^{1/(k+1)} \ell^D)$ which can be used to obtain an expression for 
the initial axial strain $\varepsilon_{z1}$ at which on average there is 
one crack per mesovolume.  

To obtain the inverse temperature $\beta_1$ corresponding to this 
initial strain,    the exact probability of observing 
a particular type of crack state is determined and  compared 
to our temperature-dependent Boltzmannian.  The particular states we choose 
to analyze are, for simplicity,  those having  
precisely one broken cell.   

The probability $p_j$ of a  state  consisting of  
 one broken cell and $N-1 $ unbroken cells  can be written       
\begin{equation}
p_j = P_1 (1-P_1)^{N-1} \Pi_{\bm x}\left[1- \delta P({\bm x})\right]
\end{equation}
where $P_1$ is again the probability of having a single broken cell and $(1-P_1)^{N-1}$ 
is the probability of having $N-1$ broken cells in the absence of other 
cracks.    
  Thus, the product 
$\Pi_{\bm x}\left[1- \delta P({\bm x})\right]
$ is the probability that no cells broke due to the strain perturbations 
caused by the presence of 
a first broken cell where ${\bm x}$ represents distance from this first broken cell. 
 We define  $\delta E_2({\bm x})$  
as the elastic energy change in a mesovolume when a second cell breaks solely 
in the perturbed strain field    
 emanating from  a first broken cell.  This energy 
varies with the separation distance $|{\bm x}|$ between the two cracks as  $|{\bm x}|^{-D}$. 
We  have
\begin{equation}
\delta P({\bm x}) = \int_{0}^{ \delta E_2({\bm x})}    \!\!
 \pi(e) \, de = \left(\frac{\delta E_2({\bm x})}{\Gamma d^{D-1}}\right)^{k+1} 
= \frac{c_2}{|{\bm x}|^{D(k+1)}} 
\end{equation}
where Eq.\ (\ref{pi}) was used for $\pi$ and where 
$c_2$ depends on both the overall applied strain and the angle from 
the first crack's orientation to the second crack.  Since $\delta P$ is small 
compared to one   
(restricting to models where cracks are smaller than the cell size $\Lambda$, since the separation distance $|{\bm x}|$ always exceeds it), we have    
\begin{equation}
\Pi_{\bm x} [1-\delta P({\bm x})] = 
1- \frac{1}{\ell^D}\int_{|{\bm x}|>\Lambda} \frac{c_2} {|{\bm x}|^{D(k+1)}} d^{D}{\bm x} 
\end{equation}
and since  $k>0$, this spatial integral over the mesovolume can be neglected in 
the thermodynamic limit.   

The conclusion is  that 
\begin{equation}
p_j = P_1(1-P_1)^{N-1} = p_0 \frac{P_1}{1-P_1} = p_0 e^{-\ln{(N-1)}} 
\end{equation}
where $p_0=(1-P_1)^N$ is the probability of the entirely intact state. 
This can be compared to our  probability law where,  
from Eqs.\ (\ref{hamiltonian})--(\ref{irrevenergy}), we have   
\begin{equation}
p_j = p_0 \exp{\left[\beta_1 \frac{(1-q)}{2}   
 {\bm \varepsilon}_1 \dd \delta {\bf C} \dd {\bm \varepsilon}_1\right]}.  
\end{equation}
Thus,  the  
 inverse temperature that holds 
when ${\bm \varepsilon}= {\bm \varepsilon}_1$ is  
\begin{equation}
\beta_1 = - \frac{\ell^D N^{1/(k+1)} \ln (N-1) }{(1-q) \Gamma d^{D-1}}. 
\end{equation}  

\subsection{Approximate approach to the  temperature}
The approach just taken in defining the initial conditions 
suggests a convenient  way of obtaining an approximate 
expression for the temperature. 

Consider ``dilute'' states $j$ where cracks do not significantly interact.  
In this case, the probability $P_m$ that any one cell has broken when the maximum 
strain tensor is at ${\bm \varepsilon}_m$ is again just the cumulative 
distribution $P_m = [\ell^D  {\bm \varepsilon}_m \dd \delta {\bf C} \dd {\bm \varepsilon}_m/
(2 \Gamma d^{D-1})]^{k+1}$.  In this case, the probability of observing a  non-interacting 
state $j$ consisting of ${\cal N}_j$ cracks  is $p_j = P_m^{{\cal N}_j} (1-P_m)^{(N-{\cal N}_j)}$ 
where we have forgone the analysis of the preceeding section  demonstrating  
 that  the unbroken-cell probabilities are negligibly  influenced by the strain 
perturbations from the ${\cal N}_j$ broken 
cells (at least for $k>0$). 
We  may write  
\begin{equation}
p_{j} 
  =  p_{0}\exp \left[ -\ln \left( \frac{1}{P_{m}}-1\right) {\mathcal{N}}_{j}\right] 
\label{eq:asympt,distrib,integree}
\end{equation}
where  \( p_{0}=(1-P_m)^{N} \) is the probability of the unbroken state.  

For such dilute states,  the Hamiltonian of Eq.\ (\ref{hamiltonian}) is written 
(with \( {\bm \varepsilon_m}={\bm \varepsilon } \))
\begin{equation}
E_{j}=\frac{1}{2} {\bm \varepsilon_m}\dd {\bf C}_0 \dd 
{\bm \varepsilon_m} -\frac{(1-q)}{2}{\bm \varepsilon_m}\dd \delta {\textbf {C}}\dd 
{\bm \varepsilon_m}\, {\mathcal{N}}_{j}
\end{equation}
so that  our probability law predicts 
\begin{equation}
\label{eq:asympt,distrib,shannon}
p_{j}=p_{0}\exp \left[ \frac{\beta(1-q)}{2}
{\bm \varepsilon}_m \dd \delta {\textbf {C}}\dd {\bm \varepsilon_m}
{\mathcal{N}}_{j}\right] .
\end{equation}
Upon using
 \( 1/P_m=[2\Gamma d^{D-1} /(\ell ^{D}{\bm \varepsilon_m}
\dd \delta {\textbf {C}}\dd {\bm \varepsilon_m}) ]^{k+1}\)
 and equating Eqs.\ (\ref{eq:asympt,distrib,shannon})
and (\ref{eq:asympt,distrib,integree}), the temperature is identified   
\begin{equation}
\label{temp}
\beta({\bm \varepsilon_m})
=\frac{- 2 \ln \left\{ [2\Gamma d^{D-1} /\left( \ell ^{D}{\bm \varepsilon _m}
\dd \delta {\textbf {C}}\dd {\bm \varepsilon _m}\right)]^{k+1} -1\right\} }
 {(1-q) {\bm \varepsilon_m}\dd \delta {\textbf {C}}\dd {\bm \varepsilon _m}}.
\end{equation}
 This expression for \( \beta \) 
has the expected behavior that \( \beta=-\infty \) when \( {\bm \varepsilon }_m=0 \), and 
that \( \beta \)
is a negative and  increasing function of \( {\bm \varepsilon}_m \) up to 
the strain point $P_m=1/2$ where it smoothly goes to zero.  For $P_m > 1/2$,  
 $\beta$ is a positive and increasing function of \( {\bm \varepsilon}_m \).     
Our probability law  with $\beta$ negative 
predicts the intact state to  have the greatest probability,  
while when  \( P_{m}  > 1/2 \) and $\beta$ is positive,  
 the most probable state jumps  to every
cell being broken. Although such a phase transition occurs in  fiber bundles \cite{PT02},  
we demonstrate in   
 Paper III using the exact differential equation for temperature, 
that  the localization transition always occurs prior to this 
divergent-temperature 
transition.      

We emphasize that  Eq.\  (\ref{temp}) is an approximation  to the extent that    
  due to the long-range nature
of elastic interactions, one  can never truly define a non-interacting state.
We  use it to obtain an order-of-magnitude idea of  the temperature at a given strain.  
But it should always be considered preferable to obtain the temperature by integrating 
the exact Eq.\ (\ref{tempde}) from the first-crack (or other exact) initial conditions.

\section{CONCLUSIONS}

The present theory of fracture in disordered solids works from the postulate
that the probability \( p_{j} \) of observing a mesovolume in a given emergent-crack
state \( j \) and at  a  given applied strain  can
be determined by maximizing Shannon's measure of the emergent-crack disorder
subject to constraints that come from the energy balance of brittle fracture.
These constraints are what allow non-uniform probability distributions to occur.
The validity of this postulate can be demonstrated in simpler cases \cite{PT02}
by integrating the probability distribution through history, but its general
validity in the case of rocks with interacting cracks  remains an open problem.
Our approach to answering this question is to use the statistical mechanics
that follows from our maximal-disorder postulate to make predictions about the
physical properties of real systems and to compare such predictions to laboratory
data. 

\begin{acknowledgments}
The authors  thank S.\ Roux and M.\ Holschneider for useful discussions in the early stages 
of this work,  and D.\ Lockner for sharing both his  data and knowledge of the fracture process. 
R.T.\ received financial support from the TMR network "Fractal Structures and Selforganization" 
through  EEC grant FMRXCT980183.

\end{acknowledgments}

\bibliographystyle{apsrev}

\end{document}